\documentclass[useAMS,usenatbib]{mn2e}

\usepackage{graphicx}
\usepackage{amssymb}
\usepackage{amsmath}
\usepackage{datetime}
\usepackage{subfigure}
\usepackage{soul}
\usepackage{color}
\usepackage{hyperref}

\definecolor{outline}{rgb}{1,0,0}
\definecolor{vero}{rgb}{1,0.24,0.59}

\begin{document}

\title[X-ray suppression by thin-shell instability]
{Suppression of X-rays from radiative shocks by their  thin-shell instability}

 \author[N. D. Kee et al.]
{Nathaniel Dylan Kee$^1$\thanks{Email: dkee@udel.edu},
 Stanley Owocki$^1$,
 Asif ud-Doula$^2$, \\
 $^1$ Department of Physics and Astronomy, Bartol Research Institute
 University of Delaware, Newark, DE 19716, USA\\
 $^2$ Penn State Worthington Scranton, Dunmore, PA 18512, USA.
}

\def\<<{{\ll}}
\def\>>{{\gg}}
\def\wig{{\sim}}
\def\spose#1{\hbox to 0pt{#1\hss}}
\def\ltwig{\mathrel{\spose{\lower 3pt\hbox{$\mathchar"218$}}
     R_{\rm A}ise 2.0pt\hbox{$\mathchar"13C$}}}
\def\gtwig{\mathrel{\spose{\lower 3pt\hbox{$\mathchar"218$}}
     R_{\rm A}ise 2.0pt\hbox{$\mathchar"13E$}}}
\def\+/-{{\pm}}
\def\=={{\equiv}}
\def\mubar{{\bar \mu}}
\def\mustar{\mu_{\ast}}
\def\Lambar{{\bar \Lambda}}
\def\Rstar{R_{\ast}}
\def\Mstar{M_{\ast}}
\def\Lstar{L_{\ast}}
\def\Tstar{T_{\ast}}
\def\gstar{g_{\ast}}
\def\vth{v_{th}}
\def\grad{g_{rad}}
\def\glines{g_{lines}}
\def\Mdot{\dot M}
\def\mdot{\dot m}
\def\yr{{\rm yr}}
\def\ksec{{\rm ksec}}
\def\kms{{\rm km/s}}
\def\qad{\dot q_{ad}}
\def\qlines{\dot q_{lines}}
\def\solar{\odot}
\def\Msun{M_{\solar}}
\def\msbyr{\Msun/\yr}
\def\Rsun{R_{\solar}}
\def\Lsun{L_{\solar}}
\def\Be{{\rm Be}}
\def\Rpole{R_{p}}
\def\Req{R_{eq}}
\def\Rmin{R_{min}}
\def\Rmax{R_{max}}
\def\Rstag{R_{stag}}
\def\vinf{V_\infty}
\def\Vrot{V_{rot}}
\def\Vcrit{V_{crit}}
\def\half{{1 \over 2}}
\newcommand{\beq}{\begin{equation}}
\newcommand{\eeq}{\end{equation}}
\newcommand{\beqa}{\begin{eqnarray}}
\newcommand{\eeqa}{\end{eqnarray}}
\def\phip{{\phi'}}

\maketitle


\begin{abstract}
We examine X-rays from radiatively cooled shocks, focusing on how their thin-shell instability reduces X-ray emission.
{
For 2D simulations of collision between equal expanding winds, we carry out a parameter study of such instability as a function of the ratio of radiative vs. adiabatic-expansion cooling lengths. In the adiabatic regime, the extended cooling layer suppresses instability, leading to planar shock compression with X-ray luminosity that follows closely the expected ($L_x \sim \Mdot^2$) quadratic scaling with mass-loss rate $\Mdot$.
In the strongly radiative limit, the X-ray emission now follows an expected  {\em linear} scaling with mass loss ($L_x \sim \Mdot$), but the instability deforms the shock compression into
extended {\em shear} layers with oblique shocks along fingers of cooled, dense material. The spatial dispersion of shock thermalization limits strong X-ray emission to the tips and troughs of the fingers, and so reduces  the X-ray emission (here by about a factor 1/50)
below what is expected from analytic radiative-shock models without unstable structure.
Between these two limits, X-ray emission can switch between a high-state associated with extended shock compression, and a low-state characterized by extensive shear.}
Further study is needed to clarify the origin of this ``shear mixing reduction factor'' in X-ray emission, and its dependence on parameters like the shock Mach number.
\end{abstract}


\section{Introduction}
\label{sec:intro}

Shocks that arise from collision between highly supersonic flows are a common source of X-ray emission from astrophysical plasmas. 
A prominent example is the case of colliding wind binaries (CWB's), wherein the collision is between strong, highly supersonic  stellar winds from the individual components of a massive-star binary system.
In relatively wide, long-period binaries, wind material that is shock-heated by the collision cools gradually by adiabatic expansion, leading to a spatially extended region of hot post-shock flow, with X-ray emission that is readily computed from the local density-squared emission measure.
This gives an overall X-ray luminosity that scales with the square of the mass loss rate ($L_X \sim \Mdot^2$) of the source stellar wind, and with the inverse of the distance $d$ from the star to the interaction front.
Numerical hydrodynamics simulations of such wide CWB systems have thus been quite successful in modelling both the level and, in the case of eccentric systems, the orbital phase variation, of observed X-rays from long-period CWB's such as $\eta$~Carinae and WR~140 
\citep{ParPit2009, ParPit2011, Rus2013}.

In closer, short-period binaries, the higher density at the interaction front means that cooling in the post-shock region can become dominated by  radiative emission, with an associated radiative cooling length $\ell_c \ll d$ that can be much smaller than the star to interaction-front distance $d$ that characterizes the scale for adiabatic expansion cooling \citep{SteBlo1992, Pit2009}.
Since the X-ray emission over the cooling layer is limited by the incoming kinetic energy flux, the X-ray luminosity from such radiative shocks is expected now to scale {\em linearly} with the mass loss rate, $L_X \sim \Mdot$
\citep{OwoSun2013}.

However, analytic stability analyses \citep{Vis1994} show that the narrowness of such radiatively cooled shock layers makes them subject to a non-linear {\em thin-shell instability}, wherein lateral perturbation of the interaction front causes material to be diverted from convex to concave regions, converting the direct compression from the oppositely directed flows into multiple elongated regions of strong shear.
Beginning with the pioneering work by \citet{SteBlo1992}, all numerical hydrodynamics simulations of flow collisions \citep[e.g.,][]{WalFol1998, Pit2009, ParPit2010} 
in this limit of radiatively cooled shocks indeed show the interaction front to be dominated by highly complex regions of dense, cooled gas, with little high-temperature material to emit X-rays.

While it is clear that such thin-shell instability structure is likely to reduce the X-ray emission from what is expected from a simple laminar compression analysis, so far there have been only limited attempts to quantify the level of this reduction, and how it might scale with physical, and even numerical, parameters.
\citet{ParPit2010} have emphasized the potential role of ``numerical conduction'', and other numerical effects associated with limited grid resolution, in lowering  the temperature of shock-heated regions, and so reducing the X-ray emission.
{\citet{LamFro2011} also assess the spatial grid needed to resolve the instability, and identify a related ``transverse acceleration instability'' that can further contribute to flow structure.}
Numerical resolution is certainly a challenge for simulating the small-scale structure that arises with a small cooling length $\ell_c$, especially when carried out over the much larger separation scale $d$; but even in simple planar slab collision models with grids set to well resolve this cooling length (see \S \ref{sec:advect_rhoT}), there is a substantial reduction in X-ray emission.
While numerical diffusion and other artefacts may play a part in this, it seems that much or even most of this reduction stems from a robust {\em physical} effect, namely the conversion of direct compressive shocks to highly oblique shocks along the elongated ``fingers'' of shear from the oppositely directed flow.

The simulations and analysis in this paper aim to quantify such effects of thin-shell instability-generated structure in reducing X-ray emission from radiative shocks.
{To focus on thin-shell structure, we ignore the added complexity of fully 3D models explored by other authors \citep[e.g.,][]{vanKep2011, ParGos2011} within the specific context of CWB's, using instead the simplest possible 2D simulations of equal colliding flows that still allow full development of thin-shell structure.}
Beyond CWB's, this has relevance for interpreting X-rays from shocks in other dense stellar outflows, for example the ``embedded wind shocks'' that originate from instabilities in the line-driving of hot-star winds. To explain the observed linear $L_X \sim L_{bol}$ relation between X-ray and stellar bolometric luminosity of single O-type stars,  \citet{OwoSun2013} proposed that thin-shell mixing in these radiative shocks reduces their X-ray emission in proportion to some power -- dubbed the mixing exponent -- of their mass loss rate.
The complexity of treating the non-local radiation transport {makes} it difficult to develop multi-D simulations of the structure arising from this line-driving instability, and so a general goal here is to use a study of direct shocks from opposing supersonic flows as a first test of this proposed scaling for thin-shell-mixing effects.

\begin{figure}
\includegraphics[width=0.5\textwidth]{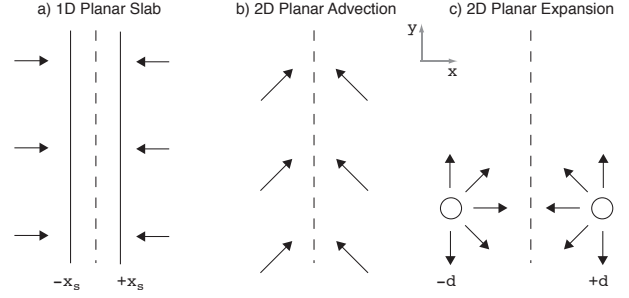}
\caption{
Schematic illustration of the 3 types of flow collision modelled here, as denoted by the figure labels. The vectors represent flow velocity, and the dashed lines represent the mean contact interface for momentum balance.  The solid lines in the 1D slab model {a} represent the shock pair located at offsets $\pm x_s$ from the interface.
The tilted vectors in the 2D planar advection model {b} indicate vertical advection at a speed equal to that of the horizontal compression.
The circles in the 2D planar expansion model {c} indicate stellar wind mass sources at locations $\pm d$ from the interface, with constant outward expansion speed.
}
\label{fig:3configs}
\end{figure}

To provide a firm physical basis, we do this through 3 tiers of simulation, based on the 3 configurations of flow collision illustrated in figure \ref{fig:3configs}.
For the simple case of a 1D planar slab with collision between equal and opposite flows (figure \ref{fig:3configs}a), we first derive analytic scalings for the temperature variation and X-ray emission (\S \ref{sec:cool_anal}) within a cooling length $\ell_c$ on each side of the interaction front, under the idealization of steady-state, standing shocks.
We next (\S \ref{sec:cool_osc_rhoT}) use time-dependent numerical simulations to illustrate the cooling oscillation \citep{CheIma1982} of such 1D slab collision, along with associated variation in X-ray emission.
\S \ref{sec:advect_rhoT} extends this planar collision model to include vertical advection in 2D (figure \ref{fig:3configs}b), showing how the initial cooling oscillation breaks up into extensive shear structure along the vertical advection, with an associated factor $\sim1/50$ reduction in the X-ray emission.
For 2D CWB-like models of collision between mass sources with planar expansion (figure \ref{fig:3configs}c), \S \ref{sec:source} carries out a systematic parameter study of how the structure formation, and X-ray reduction, depend on the mass source rate and an associated 2D cooling parameter $\chi_{2D} \sim 1/\Mdot$.
A key result is that $L_X$ follows the expected analytic scalings for the adiabatic regime $\chi_{2D} \gtrsim 1$, but is reduced by a nearly fixed factor $\sim1/50$ from the linear $L_X \sim \Mdot$ scaling expected for the strongly radiative limit $\chi_{2D} \ll 1$.
The final section (\S \ref{sec:summary}) discusses open issues from these simulations and outlines directions for future work.

\section{Cooling Analysis}
\label{sec:cool_anal}

\subsection{General Equations of Hydrodynamics}
\label{sec:genhydro}

All  flow models in this paper assume no gravity or other external forces, so that the total advective acceleration in velocity ${\bf v }$ stems only from gradients in the gas pressure $p$,
\beq
 \frac{D\mathbf{v}}{Dt} \equiv  \frac{\partial \mathbf{v}}{\partial t} +  \mathbf{v} \cdot \nabla  \mathbf{v}  =-\frac{\nabla p }{\rho}
 \, .
\label{eqn:pcons}
\eeq
Here the mass density $\rho$ satisfies the conservation condition,
\beq
 \frac{\partial \rho}{\partial t} +  \nabla  \cdot (\rho \mathbf{v} ) = 0 \, .
 \label{eqn:mcons}
 \eeq
The internal energy density, given by $e = (3/2) p$ for a monatomic ideal gas, follows a similar conservation form,
but now with non-zero terms on the right-hand-side to account for the sources and sinks of energy,
\beq
 \frac{\partial e}{\partial t} +  \nabla  \cdot (e \mathbf{v} ) = 
- p  \nabla\cdot \mathbf{v} - C_{rad}
  \, .
  \label{eqn:econs}
 \eeq
 Here the pressure term represents the effect of compressive heating ($\nabla \cdot {\bf v} < 0$) or expansive cooling ($\nabla \cdot {\bf v} >  0$), and the $C_{rad}$ term accounts radiative cooling\footnote{For the shock-heated flows considered here, we do not explicitly include any external heating term; but to mimic the effect of stellar photoionization heating in keeping circumstellar material from falling below a typical hot-star effective temperature \citep{Dre1989}, we do impose a `floor' temperature $T=30,000$~K.}.
This volume cooling rate has the scaling,
\beq
C_{rad} = n_e n_p \Lambda (T)  =   
\rho^2 \Lambda_m (T)
\, ,
\label{eq:qrad}
\eeq
where $\Lambda (T)$ is the optically thin radiative loss function
\citep{CooChe1989,SchKos2009},
 and the latter equality defines a mass-weighted form $\Lambda_m \equiv \Lambda/\mu_e \mu_p$.  
For a fully ionized plasma the proton and electron number densities $n_p$ and $n_e$ are related to the mass density $\rho$ through the associated hydrogen mass fraction $X= m_p/\mu_p = m_p n_p/\rho  $ and mean mass per electron  $\mu_e = \rho/n_e = 2 m_p/ (1+ X)$. We assume here the standard solar hydrogen abundance $X=0.72$.
For all numerical simulations below, we use the radiative loss tabulation from \citet{CooChe1989}, implemented within the exact integration scheme from \citet{Tow2009}. 

Using the ideal gas law $p = \rho kT/\bar{\mu}$ (with $k$ the Boltzmann constant and $\bar{\mu} = 0.62 m_p$ the mean atomic weight), we can combine eqns.\ (\ref{eqn:mcons}) and (\ref{eqn:econs}) to derive a general equation for the total advective variation of the temperature,
\beq
\frac{1}{T}  \frac{DT}{Dt}  = - \frac{2}{3} \, \nabla \cdot \mathbf{v} - \frac{2}{3} \frac{\bar{\mu} \rho \Lambda_m(T)}{kT}
\, .
\label{eqn:DTDt}
\eeq

\subsection{Planar steady shock with cooling length $\ell_c$}
\label{sec:lcool}

To provide a basis for interpreting the time-dependent numerical models below, let us first consider the  idealized case
in which two highly supersonic, {\em planar} flows with the same fixed density $\rho_o$ and equal but opposite speeds $v_o$ along the $x$-direction collide at a fixed interface position $x=0$, resulting in a pair of standing, {\em steady-state} shocks at  fixed positions $x=\pm x_s$ on each side of the interface (see figure \ref{fig:3configs}a).
Since  $\rho v$ is constant,  
eqn.\ (\ref{eqn:DTDt}) for temperature variation in the post-shock layers within $|x| < x_s$  takes the form
\begin{equation}
-\frac{v}{T} \frac{dT}{dx}=\frac{2}{3}\frac{dv}{dx}+\frac{2}{3}\frac{\bar{\mu}\rho\Lambda_m(T)}{kT} \, .
\label{eqn:1D_steady_dT}
\end{equation}
Because the post-shock flow is subsonic, the final deceleration to zero speed at the interface can be achieved with only a mild gradient in pressure; thus, with only a minor fractional correction in the energy balance ($\sim 1/16$; see \citet{AntOwo2004} and \S \ref{sec:fx1dslab} below), the cooling layer can be approximated as {\em isobaric}.
Together with constancy of the mass flux $\rho v$, this means $p \sim \rho T  \sim T/v $ are all constant, allowing us to eliminate density and velocity variations in terms of fixed post-shock values,
\begin{equation}
-T^2\frac{dT}{dx}=\frac{2}{5}\frac{\bar{\mu}(\rho_s T_s)^2\Lambda_m}{k(\rho_s v_s)}
\, ,
\label{eqn:dTdx}
\end{equation}
where these post-shock (subscript ``$s$") values are set by the
strong-shock jump conditions\footnote{These jump conditions are most generally cast in terms of the incoming speed {\em relative to the shock}, but in the present idealization of a steady {\em standing} shock this is just set by the speed $v_o$ measured relative to the fixed interface.},
\begin{align}
\label{eqn:RHjump}
\begin{split}
v_s&=\frac{v_o}{4} \\ 
\rho_s &= 4\rho_o  \\
T_s&=\frac{3}{16} \frac{\bar{\mu}}{k} \, v_o^2
= 14 \, {\rm MK} \, 
v_8^2 
 \,.
\end{split}
\end{align}
The latter evaluation for post-shock temperature again assumes a fully ionized gas with solar metallicity, 
with 
$v_8 \equiv v_o/10^8$cm\,s$^{-1}$.

For such typical post-shock temperatures $T_s \gtrsim 10$\,MK, $\Lambda_m$ is a quite weak function of temperature.
If we thus make the further simplification that this $\Lambda_m$ is strictly constant, 
direct integration of eqn.\ (\ref{eqn:dTdx}) yields an analytic solution for the decline in temperature from the post-shock value $T(x_s)=T_s$ to a negligibly small value at the interface ($x=0$),
\begin{equation}
\left ( \frac{T(x)}{T_s} \right )^3 = \frac{|x|}{\ell_c} ~~~ ; ~~~ |x| \le x_s=\ell_c 
\, ,
\label{eqn:Tofx}
\end{equation}
where the total cooling length $\ell_c = x_s$  from the shock to the contact interface has the scaling,
\begin{equation}
\ell_c \equiv \frac{5}{6}\frac{k v_s T_s}{\bar{\mu} \rho_s \Lambda_m}
= \frac{5}{512}\frac{v_o^3}{\rho_o \Lambda_m}
\approx 1.4 \times 10^{11} {\rm cm} \, \frac{v_8^3}{\rho_{-14}}
\, .
\label{eqn:cooling_length}
\end{equation}
{At} temperatures of the order of $10$\, MK, the radiative loss function has a value $\Lambda = \mu_e \mu_p \Lambda_m \approx 3 \times 10^{-23}$\,erg\,cm$^3$\,s$^{-1}$ \citep{CooChe1989,SchKos2009}, giving the numerical scalings in the last equality,  with $\rho_{-14} \equiv \rho_o/10^{-14}$\,g\,cm$^{-3}$.

We can also define a characteristic post-shock cooling time as the post-shock flow speed through this cooling length,
\beq
t_c  \equiv \frac{\ell_c}{v_s} =  1.2 \times 10^4  {\rm s} \, \frac{v_8^2}{\rho_{-14}}
\, .
\label{eqn:tauc}
\eeq
The values for $\ell_c$ and $t_c$ provide convenient reference scales for the cooling length and time in the more general numerical models below.

\begin{figure*}
\includegraphics[width=1\textwidth]{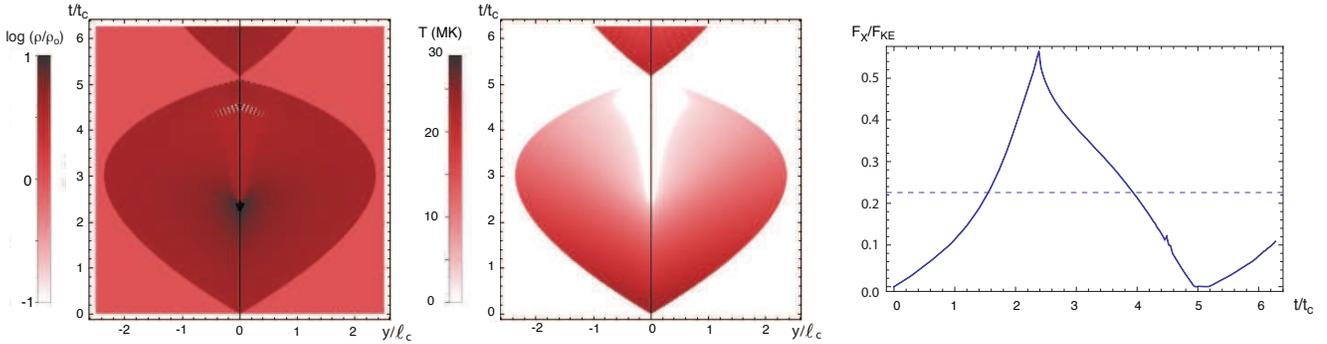}
\caption{The 1D cooling oscillation in  density $\log( \rho (x,t)/\rho_o)$ (left-hand panel) and temperature $T(x,t)$ (in MK; middle panel), plotted vs.\ horizontal position $x$ (in units of the cooling length $\ell_c$) and time $t$ (in units of the cooling time $t_c$). The right-hand panel shows the resulting time variation of the X-ray flux $F_X$, in units of the total flow energy flux $F_{KE} = \rho_o v_o^3$, with the horizontal dashed line showing the time-averaged value over a cooling oscillation cycle.
}
\label{fig:1D_cool_osc}
\end{figure*}

\subsection{X-ray flux from planar, steady shock}
\label{sec:fx1dslab}

This simple model of steady shock in a 1D slab also provides a useful illustration for the scaling of the X-ray emission. The volume emissivity for radiation of energy $E$ depends on the local density and temperature,
\beq
\eta(E,x) = \frac{\rho^2 (x) }{\mu_e \mu_p} \Lambda(E,T(x)) 
\, ,
\label{eqn:etaE}
\eeq
where the energy-dependent radiative loss function  $\Lambda(E,T)$ (erg\,cm$^3$\,s$^{-1}$\,keV$^{-1}$) is derived from APEC thermal equilibrium emission models \citep{SmiBri2001}, with the total radiative loss function $\Lambda(T) = \int_0^\infty \Lambda(E,T) \, dE$.
The associated radiative flux comes from integration over both shock cooling layers,
\beqa
F(E) &=& \int_{-x_s}^{x_s} \frac{\rho^2 (x) }{\mu_e \mu_p}  \Lambda(E,T(x)) \, dx 
\label{eqn:FEint}
\\
&=& \frac{15}{16} \, 
\rho_o v_o^3
\int_0^{T_s} \frac{\Lambda(E,T)}{\Lambda(T)} \, \frac{dT}{T_s}
\, ,
\label{eqn:FE}
\eeqa
where the latter equality comes from using  eqn.\ (\ref{eqn:dTdx}) to change integration variable to temperature, and the factor $15/16$ reflects the above-mentioned 1/16 loss due to neglect of compressive work within the isobaric cooling model
\citep{AntOwo2004}.
The total flux above some representative X-ray threshold, taken here to be $E_X = 0.3$\,keV, thus just scales with the kinetic energy flux $F_{KE} \equiv \rho_o v_o^3$ from both sides,
\beq
F_X = \int_{E_X}^\infty F(E) \, dE 
\approx 
F_{KE} \, f_X 
\, ,
\label{eqn:FXslab}
\eeq
where  
$f_X$ characterizes the fraction of total radiative loss that is emitted in the X-ray bandpass above $E_X$. Note that for  $E_X =0$, we have $f_X=1$, so that, apart from the 1/16 loss,  the bolometric radiative flux nearly equals the total incoming kinetic energy flux.

An important point to emphasize here is that, even though the local volume emissivity (\ref{eqn:etaE}) scales with {\em density-squared emission measure}, the integrated fluxes (\ref{eqn:FE}) and (\ref{eqn:FXslab}) over such a radiative cooling layer scale only {\em linearly} with the inflow density $\rho_o$.

\section{Numerical simulations
for laminar flow collisions}
\label{sec:lamflow}

\subsection{The 1D cooling oscillation}
\label{sec:cool_osc_rhoT}

While the above analytic solution for a presumed steady-state shock provides a simple overall characterization of  the shock cooling layer, the linear stability analysis of \cite{CheIma1982} shows that, even for a 1D steady laminar incoming flow, a post-shock layer undergoing such radiative cooling is generally {\em not} steady-state, but is {\em unstable to cooling oscillation} modes.

As a basis for 2D models below, let us first examine 1D numerical simulations of this cooling oscillation for this case of direct collision of two equal and opposite laminar flows.
Using the PPM \citep[Piecewise Parabolic Method;][]{ColWoo1984} numerical hydrodynamics code 
VH-1\footnote{http://wonka.physics.ncsu.edu/pub/VH-1/},
we solve the time-dependent conservation equations (\ref{eqn:pcons}) --
(\ref{eqn:econs}) on a fixed 1D planar grid of $n_x = 1000$ zones extending $\pm 2.5 \ell_c$ on each side of central contact discontinuity (again set at $x=0$) between the two flows; the uniform spatial zones of size $\Delta x= 0.005 \, \ell_c$ thus very well resolve the cooling region.
The left and right boundaries at $x \approx \pm 2.5 \ell_c$ assume highly supersonic inflow at speeds $v_o = 1000$\,km\,s$^{-1}$ of  gas with a typical hot-star temperature $T=30,000$\,K and thus sound speed $a \equiv \sqrt{kT/\bar{\mu}} \approx 20$\,km\,s$^{-1} = v_o/50$.
The initial condition at $t=0$ extends these inflow conditions to direct collision at the $x=0$ contact surface.
Since in practice stellar photoionization heating tends to keep gas from cooling much below the stellar effective temperature, we also use $T_o$ as a ``floor'' temperature for both pre- and post-shock gas \citep{Dre1989}.

The color scale plots in figure \ref{fig:1D_cool_osc} show the time evolution (plotted along {the} vertical axis in units of  the cooling time $t_c$) of the density  and temperature within the cooling layers on each side of the contact discontinuity (vertical dark line), bounded on the outside edges by the oscillating shock discontinuity.  
Initially, the limited factor 4 compression means, from mass continuity, that the shock location propagates back from  the contact at a speed $v_o/3$.
However, as the gas near the contact cools, the lower pressure allows compression to higher density, which slows and then reverses this back propagation, but with a roughly factor two {\em overshoot} of the equilibrium cooling length.  As the gas near the contact continues to cool, the cooling  layer contraction leads to direct collapse of the shock on to the contact, whereupon the cycle repeats, leaving just a buildup of cold dense material at the contact discontinuity.  
The overall period of the oscillation is a few ($\sim$5) cooling times, with an amplitude of a couple cooling lengths.

During the initial phase of the oscillation, the back-propagation of the shock means the net velocity jump in the shock frame actually exceeds, by a factor 4/3, that in the steady flow model, and so this leads to an initial post-shock temperature that is $(4/3)^2 = 16/9$ higher than given in the steady shock scalings of eqn.\ (\ref{eqn:RHjump}).
On the other hand, during the contraction phase, the shock velocity jumps are weaker, leading to lower post-shock temperatures.

This 1D model of the cooling oscillation provides another sample case for deriving  X-ray emission from such shock compressions.
For each time $t$ of a simulation,  we carry out the 1D integration (\ref{eqn:FEint}) to obtain the now time-dependent flux at a selected energy, $F(E,t)$.
Integration of $F(E,t)$ over an X-ray energy bandpass $E_X > 0.3$\,keV yields the associated X-ray flux $F_X (t)$. 
For this cooling oscillation model, the rightmost panel of figure \ref{fig:1D_cool_osc} plots the time variation of $F_X$, normalized by the total
kinetic energy flux from both sides of the inflow, $F_{KE} = \rho_o v_o^3$.
Note that{, due to a accumulation of shock heated material,} $F_X$ increases up to the time $t \approx 3 t_c$ when the cooling region reaches its maximum size, with a peak that
reaches about half the kinetic energy input rate.
But as noted above, during the subsequent cooling zone compression the weaker shocks give lower temperatures. This abruptly reverses the $F_X$ into a decline, making it nearly vanish near the minimum.

\begin{figure}
\includegraphics[width=0.5\textwidth]{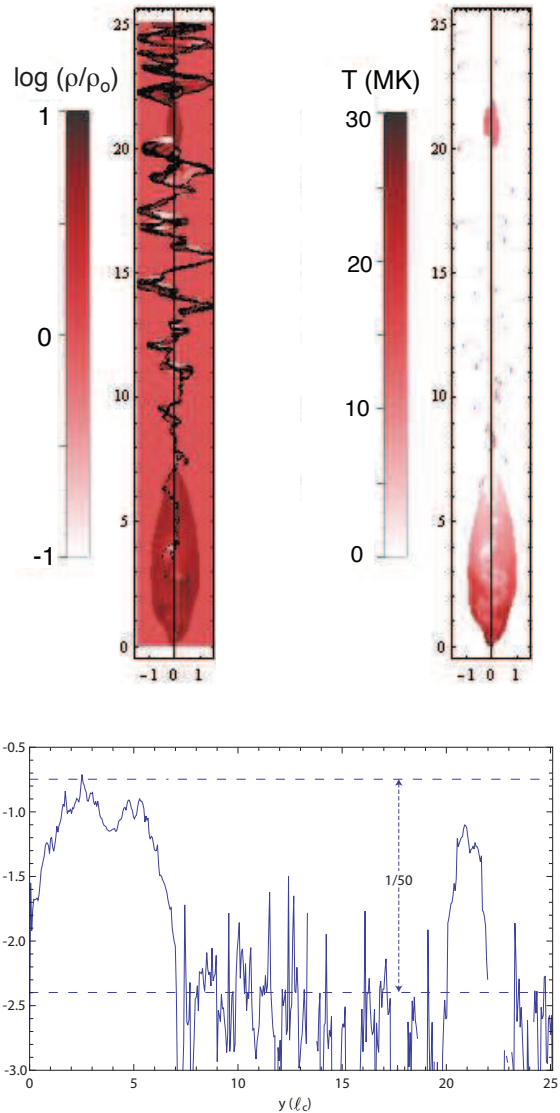}
\caption{Evolved-time snapshots {of the} 2D advection model showing spatial $(x,y)$ variations of log density (upper left) and  temperature (upper right), along with vertical {\bf $(y)$} variation of $x$-integrated X-ray emission $F_X (y)$ (bottom: now on a log scale, but again normalized by the total kinetic energy flux $F_{KE} = \rho_o v_o^3$). 
The horizontal dashed lines compare the $F_X$ time-averaged compressive value from the 1D cooling oscillation (upper)  and the final  shear-dominated state (lower), showing a roughly factor 1/50 reduction in the X-ray emission.}
\label{fig:advectFrames}
\end{figure}

\subsection{2D spatial breakup of density and temperature structure}
\label{sec:advect_rhoT}

Let us next consider a 2D simulation of this same basic model of direct collision between two equal and opposite planar flows.
In addition to the temporal variation from the cooling oscillation, the addition of a second, {\it transverse} (y) direction now  allows the possibility of a {\em spatial} break-up of  the post-shock cooling layer.
To illustrate this within a single time snapshot, it is convenient to introduce a constant {\em vertical advection} speed $v_y$, set here to be the same as the horizontal (x) inflow compression speed $v_o$, so that material introduced at all boundaries enters the computational domain at a fixed angles of $\pm 45^\circ$ on each side of the central axis $x=0$.
At the lower boundary $y=0$, this leads to a direct interaction at this $x=0$ interface. {The flow also makes a 45$^\circ$ angle with this interface, but because the shock normal speed is kept at the same $v_o=v_x=1000$~km s$^{-1}$ as in the 1D model, the shock strength is the same. The advection thus represents a simple Galilean transformation of what would occur in a 2D extension of the 1D direct collision model. Moreover, away from the lower boundary, it provides a convenient way to visualize the temporal evolution of the structure through a single time snapshot.}


Figure \ref{fig:advectFrames} plots such a snapshot of the density (upper left) and temperature (upper right) structure at a time $t \approx 12 t_c$, corresponding to twice the advection time {\bf$t_{adv}=y_{max}/v_0$} through the vertical extent {\bf$y_{max}$}.
Both spatial directions are now scaled by the cooling length $\ell_c$, but for the vertical axis the simple advection at a fixed speed means that each scaled length unit can be readily translated to  time in cooling times via 
$y/\ell_c = (4 v_y/v_o) (t/t_c) = 4 (t/t_c)$; this thus allows for direct comparison with the 1D space + time variation plot in figure \ref{fig:1D_cool_osc}.
The computational grid now contains $n_x=600$ horizontal zones over the range $-1.5 \ell_c < x  < +1.5 \ell_c$, and $n_y=5000$ vertical zone over the range $0 < y < 25  \ell_c$. This agains corresponds to a uniform mesh size  $\Delta x = \Delta y  = 0.005 \, \ell_c = \ell_c/200$ that  very well resolves the 1D cooling length.

Note that the cooling oscillation still appears in the initial flow interaction near the lower boundary, but the dense cooled material near the central contact  quickly breaks up into a complex structure. The enhanced cooling associated with mixing of this dense structure reduces both the amplitude and period of the oscillation. Indeed, after just one initial  cycle, the 1D compressive cooling oscillation is now effectively overridden by an extensive {\em shear} structure, which grows with increasing distance (or advection time) from the lower boundary, forming complex ``fingers'' of cool, dense gas that bound regions of oppositely directed flow.

The net result is to transform the strong shock compression of the 1D collision into a complex of extended {\em shear }layers, along which any shocks are very oblique and thus very weak, with direct shocks limited to very narrow regions at the (convex) ``tips''  and (concave) ``troughs'' of the fingers.
As shown in upper right panel of figure \ref{fig:advectFrames}, this leads to a corresponding reduction in the spatial extent of high-temperature gas in the downstream flow. The bottom panel of figure \ref{fig:advectFrames} shows that, {after an initial sharp rise due to the accumulation of strong compressive shock heating near the lower boundary}, the associated horizontally integrated X-ray emission {$F_x$ drops abruptly after a single cooling oscillation extending over a few cooling lengths. It  then varies greatly with y-position,} but apart from a limited segment of more direct compression around $y \approx 20 \ell_c$, the overall level after the initial oscillation cycle is reduced by about a factor 1/50 
(as denoted by the range between the horizontal dashed lines).

\begin{figure}
\includegraphics[width=0.5\textwidth]{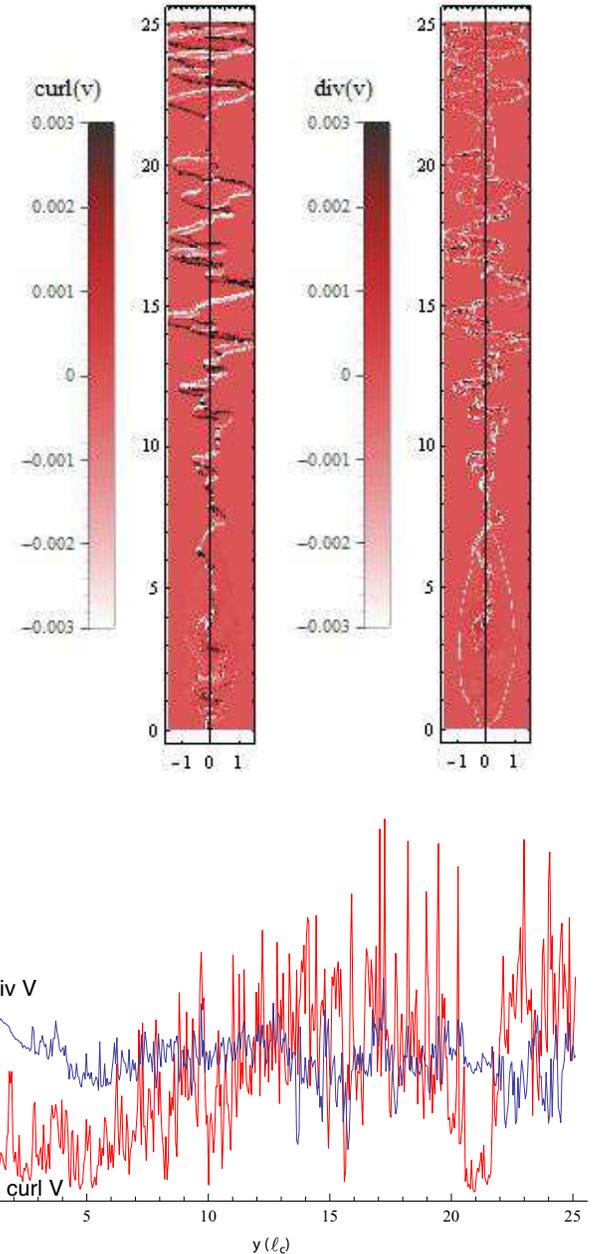}
\caption{{Snapshots of the 2D advection model at the same time as figure \ref{fig:advectFrames}, showing spatial $(x,y)$ variations of velocity divergence (upper left) and curl (upper right). The bottom panel plots vertical evolution of rms horizontal averages of these quantities. Note that from the initial condition at the lower boundary (y=0) the divergence decreases while the curl increases.}}
\label{fig:div_curl}
\end{figure}

Note that an important aspect of this thin-shell-instability reduction in X-rays is the transformation of the flow compression, characterized by a velocity divergence $\nabla \cdot {\bf v}$, into strong velocity shear, characterized a flow vorticity $\nabla \times {\bf v}$. {The spatial distribution of these quantities, as well as their relative vertical evolution, is illustrated in figure \ref{fig:div_curl}.} 

\section{Collision between Expanding Outflows}
\label{sec:source}

In practice, shock collisions in astrophysics often occur in {\em expanding} outflows, for example in the collision between two spherically expanding stellar winds in a binary system.
Such expansion tends to give the velocity divergence on the right side of the temperature equation (\ref{eqn:DTDt}) a positive value, contributing then to an {\em adiabatic expansion cooling} that competes with the radiative cooling term 
\cite[see, e.g.,][]{SteBlo1992}.
To examine how this competition affects radiative cooling instabilities and their associated reduction in shock temperature and X-ray emission, let us now generalize the above flow collision models to allow for expansion within the 2D plane.

\begin{figure*}
\includegraphics[width=1\textwidth]{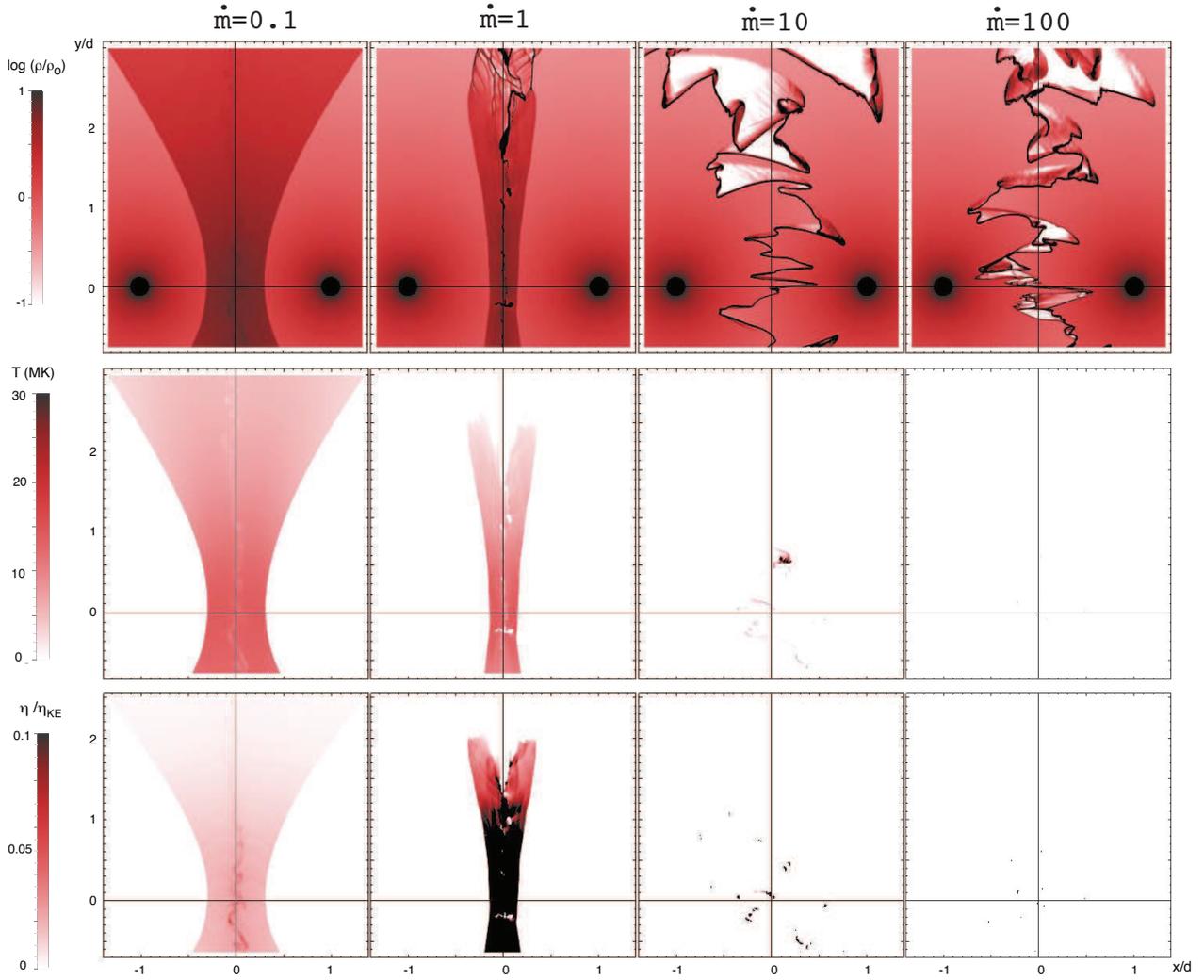}
\caption{
Final time ($t_f=32 d/v_o$) snapshots for models with mass cooling parameter $\mdot =$ 0.1, 1, 10, and 100, arranged in columns from left to right, with rows showing spatial $(x,y)$ variation of density $\log \rho$ (top),  temperature $T$ (middle), and X-ray emissivity $\eta_X$ (bottom).
The temperature is in MK, but the density is scaled by $\rho_o(d) = \Mdot/(2 \pi v_o d)$ and the emissivity by an associated kinetic power density,
 $\eta_{KE} \equiv \Mdot v_o^2/(2 \pi d^2)$. The axes for horizontal (x) and vertical (y) position are  in units of the star-interface distance $d$.
}
\label{fig:sourceFrames}
\end{figure*}

\subsection{Cooling parameter $\chi_{2D}$ for 2D models with planar expansion}
\label{sec:chi2D}

Specifically, instead of two opposing laminar flows, let us now assume the configuration in figure \ref{fig:3configs}c, namely
a 2D {\em planar expansion} from two distinct  localized mass sources at positions $\{x_m,y_m\}=\{\pm d, 0\}$, with equal constant outflow speed $v_o$ and equal mass ejection rate $\Mdot$.
To facilitate connection to 3D outflows characterized by a volume density $\rho$, we can formally assume a {\em cylindrical} expansion, with $\Mdot$ representing the mass source rate {\em per unit length} along an arbitrarily extended $z$-axis perpendicular to the 2D ({x,y}) computational plane.  Ahead of any interaction, at a distance $r = \sqrt{(x-x_m)^2+y^2}$ from either source, mass conservation for steady-state, constant-speed, cylindrical expansion implies a mass density  $\rho = \Mdot/2 \pi v_o r$.
The equality of the two mass sources means that their interaction will again center on the bisector symmetry line at $x=0$.

Along the $x$-axis line between the sources at $x_m = \pm d$, this means that the mass flux within the cooling layer now varies as {\bf $\rho v = \Mdot/2 \pi r$}.
Applying this for the isobaric post-shock cooling layer model of \S \ref{sec:lcool}, the temperature eqn.\ (\ref{eqn:dTdx}) can now be 
written in a generalized, scaled form that accounts for this $r$-dependence of the mass flux,
\begin{equation}
T^2 \frac{\partial T}{\partial r}=-\frac{2 T_s^3 }{3 \chi_{2D}} \, \frac{r}{r_s^2}
\, , 
\label{eqn:dTdrs}
\end{equation}
which applies along the ($y=0$) $x$-axis, with $r=d-|x|$.
Here the shock distance from the source is set by
\beq
r_s = \frac{d}{\sqrt{1+\chi_{2D}}}
\, ,
\label{eqn:rsanal}
\eeq
where the dimensionless cooling parameter,
\begin{equation}
\chi_{2D} \equiv \frac{5\pi v_o^4}{128\Lambda_m \Mdot}
= \frac{2 \ell_c}{d}
\, ,
\label{eqn:chi_2D}
\end{equation}
with $\ell_c$ defined here by eqn. (\ref{eqn:cooling_length}) using the inflow density at the interface, $\rho_o (d)=\Mdot/2\pi v_o d$.
Integration of eqn. (\ref{eqn:dTdrs}) with the requirements that $T(r_s)=T_s$ and $T(d)=0$ now gives for the temperature variation within the cooling zone along the $x$-axis,
 \beq
 \left ( \frac{T(r)}{T_s} \right )^3 = \frac{1}{\chi_{2D}} \, \left ( \frac{d^2-r^2}{r_s^2} \right )
 \, ; ~~~ r_s < r < d ~~ , ~~ y=0
 \, .
 \label{eqn:Trcyl}
 \eeq
 For $\chi_{2D} \ll 1$, $r_s/d \approx 1-\chi_{2D}/2 = 1 - \ell_c/d$, and since $r/d = 1- |x|/d$, we find from first-order expansion in $|x|/d \ll 1$, that eqn. (\ref{eqn:Trcyl}) recovers the laminar-collision scaling (\ref{eqn:Tofx}).
 
The dimensionless parameter $\chi_{2D}$ characterizes the ratio of length scales for cooling vs. expansion, set by $\chi_{2D} = 2 \ell_c/d$.
It serves as a 2D analogue to the commonly quoted, standard cooling parameter $\chi$, defined by  \cite{SteBlo1992} in terms of the ratio of {\em time} scales for cooling vs.\ expansion in the full 3D case of colliding stellar winds.
Both have identical scalings with the fourth power of the flow speed and inverse of the mass source rate; but note that, unlike the full 3D case,  the 2D scaling here has no dependence on the separation distance $d$.  Because planar expansion has one less dimension than the full spherical case, this 3D  scaling with distance in the numerator becomes replaced with division by a mass source  rate $\Mdot$ {\em per unit length} from cylindrical sources that formally extend perpendicularly from the 2D plane.

Nonetheless, for a fixed flow speed $v_o$, which through eqn.\ (\ref{eqn:RHjump}) sets the post-shock temperature $T_s$, we can still readily examine the effects of varying the relative importance of radiative vs.\ adiabatic cooling by adjusting this mass source rate $\Mdot$ to change the parameter $\chi_{2D}$.
Indeed, since $\chi_{2D} \sim 1/\Mdot$, it is convenient to characterize the radiative cooling efficiency in terms of $\mdot \equiv \Mdot/\Mdot_1$,  where $\Mdot_1\equiv 5 \pi v_o^4/(128 \Lambda_m)$ is the mass source rate for which $\chi_{2D} = 1$.
This gives $\mdot = 1/\chi_{2D}$.

In the strong radiative cooling limit $\mdot \gg 1$, the cooling region in this steady shock model is confined to {a} narrow layer with $|x| \le \ell_c = d/2\mdot$ on each side of the interaction front at $x=0$. 
Following the planar scaling (\ref{eqn:FXslab}), the X-ray emission in this case should increase linearly with the mass flux, $\mdot$.
But the very narrowness of this layer makes it subject to the thin-shell instabilities \citep{Vis1994} that give rise to extensive spatial structure seen in the above 2D laminar-collision models, with associated reduction in shock temperature and X-ray emission.

In contrast, in the limit of inefficient radiative emission $\mdot \ll 1$, the cooling is instead by adiabatic expansion, with a much thicker offset from the interface. The X-ray emission integrated over this extended interface now depends on the density-squared emission measure, implying a total emission that scales with $\mdot^2$.   Moreover, this extended layer  can now effectively suppress the thin-shell instability, allowing the post-shock gas to retain higher temperatures, and so an extended X-ray emission.

In the absence of instability-generated structure, \citet{OwoSun2013} proposed a simple scaling law that ``bridges" the adiabatic vs.\ radiative limits, which in the current notation can be expressed in units of twice the X-ray luminosity for the transition case $\mdot=1$,
\beq
L_X \approx \frac{\mdot^2}{1+\mdot}
\, .
\label{eqn:bridgelaw}
\eeq
A general goal here is to use numerical simulations to test how this scaling is modified by the effects of the thin-shell instability for high-density flows with $\mdot > 1$.

\subsection{Numerical simulation settings}
\label{sec:source_rhoT}

Let us now quantify these expectations with a full numerical simulation parameter study that examines how shock structure and X-ray emission in this 2D model of shock collision and planar expansion depends on this cooling efficiency parameter $\mdot$.
The  numerical model again assumes a constant source flow speed $v_o=1000$\,km\,s$^{-1}$, but now with the outflow mass sources at $\{x_m,y_m\}=\{\pm d,0\}$ {\em embedded} in a uniform spatial grid with $n_x = 1024$ zones ranging over $-1.33 d < x < + 1.33 d$ and $n_y = 1204$ zones over $-0.63 d < y < 2.5 d$. This implies fixed zone sizes $\Delta x = \Delta y = 0.0026\,d = 0.0052 \, \mdot \ell$.

The separation between the mass sources is set to a typical stellar binary separation $2d=3.1 \times 10^{12}$\,cm = 0.21\,au. All models are run to a final time $t_{f}=$500\,ks, corresponding to nearly 32 characteristic flow times $t_d=d/v_o=$15.75\,ks from the sources to the interface; this is sufficient for even material that initially collides along the source axis to flow out through either the bottom or top boundary, following its vertical pressure acceleration away from this $x$-axis. The simulation uses simple supersonic outflow boundary conditions along both horizontal and vertical edges of the computational domain. 

\begin{figure*}
\includegraphics[width=1\textwidth]{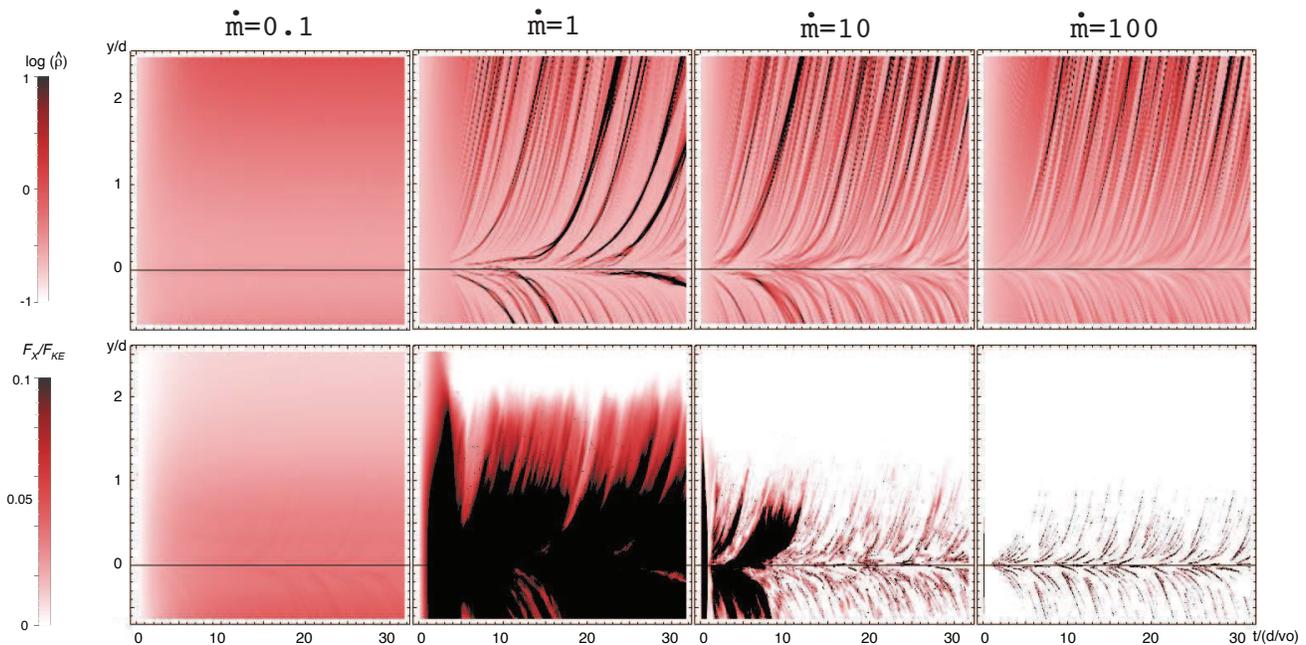}

\caption{
Time-height (t,y) variation of lateral (x) integrations of density ${\hat \rho} (y,t)$ (top row; on log scale),
for models with mass cooling parameter $\mdot =$ 0.1,  1,  10, and 100, again arranged in columns from left to right.
The X-ray flux is normalized by the total kinetic energy flux at the interaction front along the axis between the stars, $F_{KE} = \Mdot v_o^2/(2 \pi d)$.
The vertical (y) spatial axis is in units of star-interaction distance $d$, and the horizontal time axis (t) is in units of the characteristic flow time $d/v_o$.
}
\label{fig:sourceSpaceTime}
\end{figure*}

\subsection{Structure snapshots at final, evolved time}
\label{sec:source_rhoT_snap}

For the final, well-evolved time ($t = 32 d/v_o$),  figure \ref{fig:sourceFrames} compares results for simulations with $\mdot = $ 0.1, 1, 10 and 100, arranged along columns from left to right.
The upper two rows give color plots of the log density (top) and temperature (middle); the bottom row shows the associated X-ray volume emission $\eta_X$ for energies $E>E_X=3$\,keV.

The results quite vividly illustrate the trends anticipated above.

For the lowest-density case with $\mdot=0.1$ (leftmost column), the relative inefficiency of radiative vs. adiabatic cooling  leads to a distinct standoff of the shock from the interface, with only very weak instability structure forming along the interface.
Along the direct collision between the sources, the high post-shock pressure drives material away from this axis, allowing a shock standoff  radius $r_s/d \approx 0.68 $ that is larger than the $r_s/d = 1/\sqrt{1+4} = 0.45$ predicted by 
eqn.\ (\ref{eqn:rsanal}) (which accounts  for planar expansion, but not such vertical pressure acceleration).
As the standoff distance increases at larger $|y|$, there develops an extended region of high post-shock temperature, with an associated extended region of X-ray emission.

For the factor-ten higher density model with $\mdot =1$ (left central column), the cooling layers become narrower, with now quite notable instability near the interface that becomes strongly developed at large $y$. The high temperature region is accordingly narrower, with vertical extent terminated at the location that instabilities become strong.  The X-ray emission occurs over a similar spatial extent, but is now stronger in the high density regions near the $x$-axis interface.

For a further factor-ten higher-density model with $\mdot = 10$ (right central column), the cooling layers now become completely unstable, with density showing extensive finger-like structure similar to the above laminar models. The high-temperature gas, and associated X-ray emission, is now limited to small regions, again associated with the tips of the fingers and the troughs between them.

Finally, in the highest-density model with $\mdot = 100$  (rightmost column), the cooling instability leads to even more extended fingering, with the few zones of high-temperature and X-ray emission now hardly noticeable. However, as quantified below, the high density of the few remaining hot regions can still lead to significant overall X-ray luminosity

\subsection{Time-height evolution of flow structure}
\label{sec:source_rho_tvar}

To complement such single-time snapshots of the flow structure, let us next examine its time evolution.
Specifically, to illustrate the flow evolution in time $t$ and height $y$, let us define horizontal $x$-integrations of the density and X-ray emission.
For the density, to compensate for the $1/r$ decline, we weight the integration by $r$, normalized by the associated integration through the unperturbed mass source,
\beq
{\hat \rho} (y,t) \equiv   \frac{3 \pi v_o}{4 \Mdot d}  \int_{-4d/3}^{4d/3} \rho (x,y,t) \,r \, dx
\, .
\label{eqn:dyt}
\eeq
We can similarly define a horizontally integrated X-ray flux,
\beq
F_X (y,t) \equiv   \, \int_{-4d/3}^{4d/3} \eta_X (x,y,t) \, dx
\, .
\label{eqn:FXyt}
\eeq

Figure  \ref{fig:sourceSpaceTime} shows color plots of the time-height evolution of ${\hat \rho} (y,t)$ (upper row) and  $F_X (y,t)$ (lower row), again for models with $\mdot = $ 0.1, 1,  10, and 100, arranged along columns from left to right.

For the lowest-density model with $\mdot =0.1$, both density and X-ray emission quickly adjust to a time-independent steady-state with smooth distribution in $y$. But for all higher density models, there develops clear structure, with density compressions that diverge away from the mass-source $x$-axis, along with associated structure in X-ray emission.
In the $\mdot = 1$ case, and in the initial evolution of the $\mdot=10$ case,  the overall level of X-ray emission is increased in proportion to the higher source mass rate $\mdot$.

But at a time around $t \approx 13 d/v_o$, the latter case shows a sharp decline in filling fraction of X-ray emitting structures, reflecting the formation of strong fingerlike structures from the thin-shell instability, and its associated limitation of X-ray emission to tips and troughs of the fingers.
For the highest-density case $\mdot=100$ this X-ray volume reduction starts near the initial time,  and is even more pronounced.
This lower filling factor significantly reduces the X-ray luminosity, though this can be partly compensated by the more intense local emission from dense, strongly emitting regions. 

\begin{figure}
	\centering
	\includegraphics[width=0.48\textwidth]{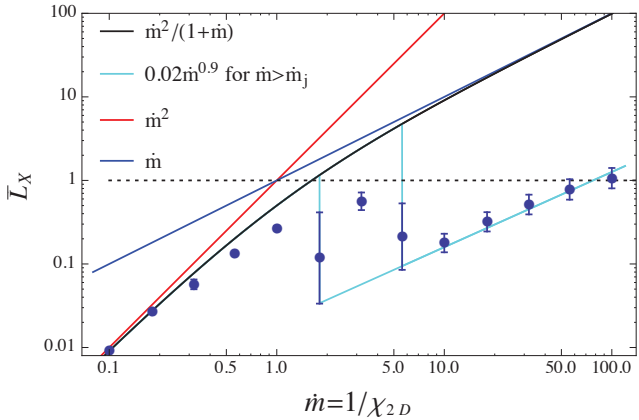}
\caption{
Time-averaged X-ray luminosity ${\bar L}_X$ vs.\ mass-loss-scaled cooling efficiency $\mdot = 1/\chi_{2D}$. 
The legend identifies curve styles for  various analytic scaling formulae. 
The points for simulation results are scaled by anchoring ${\bar L}_X$ for the lowest-density case $\mdot = 0.1$ to the bridging form (\ref{eqn:bridgelaw}),
 with error bars representing $\pm 1\sigma$ time variability.  
In the nearly adiabatic regime $\mdot \lesssim 1$, the simulations nearly follow the black curve for the scaling (\ref{eqn:bridgelaw}).
In the strongly radiative limit $\mdot \gg 1$, they are (almost) linear with $\mdot$, but with strong reduction below the analytic form (\ref{eqn:bridgelaw}).
The transition regime $2 < \mdot <  6 $  has large variations from switching between high and low ``bi-stable'' states. The cyan curves show factor 1/50 jumps at values $\mdot_j$ bracketing this transition regime,
with slightly sublinear ($\sim\mdot^{0.9}$) scaling above the jump.
}
\label{fig:masterPlot}
\end{figure}

\subsection{Scaling of X-ray luminosity with $\mdot$}
\label{sec:source_master}
Let us now quantify the overall scaling of the X-ray luminosity with the cooling efficiency $\mdot = 1/\chi_{2D}$. 
For this we derive the asymptotic time-average  X-ray luminosity ${\bar L}_X$ as the integral of $F_X (y,t)$ over positive $y$, with then a  time average over the time $t > 13 d/v_o$ after transition to its asymptotic state.  
In addition to the 4 models detailed above that differ in 1 dex increments of $\mdot$, we also compute 9 additional models to give a denser parameter grid of 13 models in 0.25 dex increments over the full range from $\mdot=0.1$ to 100.

Figure \ref{fig:masterPlot} plots (on a log-log scale) the resulting data points for $ {\bar L}_X$ vs. $ \mdot$, along with error bars to indicate the level of 1$\sigma$ temporal variation.
The black curve compares the \citet{OwoSun2013} bridging-law scaling (\ref{eqn:bridgelaw}), along with the linear $L_X \sim \mdot$ scaling (blue) and quadratic $L_X \sim \mdot^2$ scaling  (red) expected respectively in the high-density ($\mdot \gg 1$), radiative shock limit, and  in the low-density ($\mdot \ll 1$) adiabatic shock limit.
We normalize all the simulation data by anchoring the smooth emission from the lowest-density ($\mdot = 0.1$), nearly adiabatic case to exactly fit this bridging law, with 
 ${\bar L}_X = \mdot^2/(1+\mdot) = $0.009.

In figure \ref{fig:masterPlot} the first 5 data points with lowest $\mdot$, ranging from 0.1 to 1, do indeed nearly follow this simple bridging law, with small temporal variability indicated by error bars that are less than the point sizes. This is consistent with the relatively extended shock compression and X-ray emission shown in the two leftmost columns of figures \ref{fig:sourceFrames} and \ref{fig:sourceSpaceTime} for models with $\mdot$= 0.1 and $\mdot=$1. 

In contrast, for the 5 highest-density models, with $\mdot \ge 10$, the ${\bar L}_X$ all fall roughly a fixed factor 1/50 below the linear increase that applies in the $\mdot \gg 1$ limit of this bridging law, with somewhat larger, but still modest errors bars indicating only a moderate level of time variability. This is similar to the factor 1/50 reduction in X-ray emission of the 2D advection models once the initial compressive oscillation is transformed to the strong shear flow and finger structure.  It is also consistent with the extensive shear fingers and reduced X-ray emission shown in the two rightmost columns of figures \ref{fig:sourceFrames} and \ref{fig:sourceSpaceTime} for models with $\mdot$= 10 and $\mdot=$100. 

\begin{figure*}
\includegraphics[width=1\textwidth]{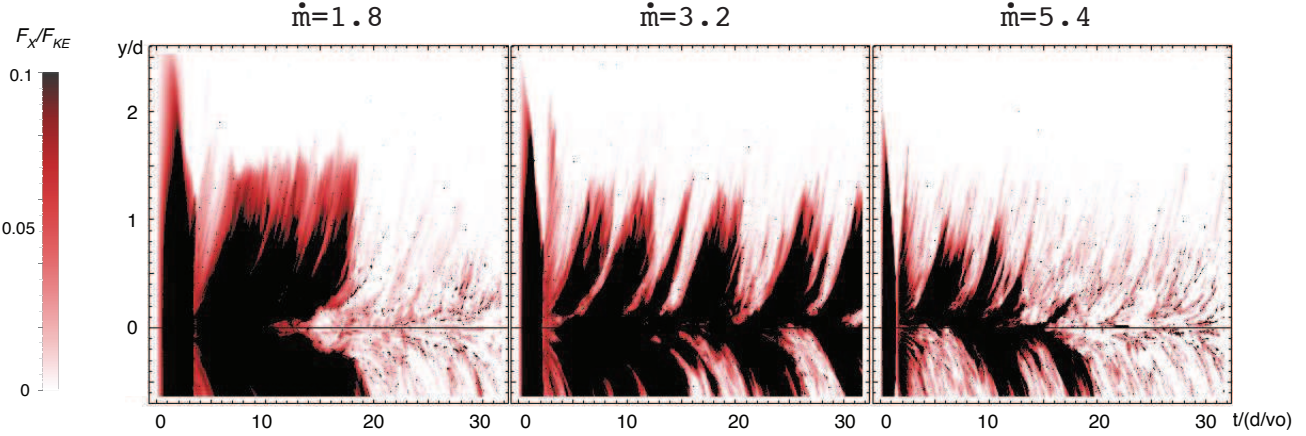}
\caption{Time-height variation of  laterally integrated X-rays for  transition cases ($\mdot = $ 1.8, 3.2 and 5.4), showing a bi-stability between relatively smooth states with high $L_X$ and highly structured states with reduced $L_X$.}
\label{fig:bistab}
\end{figure*}

Between these two limits, for the 3 intermediate cases with $ \mdot = $ 3.5, 6.3 and 5.6, the  ${\bar L}_X$ vary somewhat erratically, trending below the bridging curve, with large error bars indicating strong time variability.
Figure \ref{fig:bistab} illustrates that this apparently stems from the tendency for these intermediate cases to undergo switching between high X-ray states -- with the extended compression and emission of the low-$\mdot$, adiabatic limit --  and a low X-ray state -- with the small-scale shear and dense fingers typical of the unstable structure in the high-$\mdot$, radiative limit.

The overall result is thus that colliding-flow shock X-rays follow {closely} the simple adiabatic-to-radiative bridging law for low to moderate density shocks with $\mdot \lesssim 1$, then show a somewhat random modest declining trend for the moderately radiative cases $2 < \mdot < 8$, and finally again increase linearly with $\mdot$ for the strongly radiative cases $\mdot > 10$, but at about 1/50 lower level than the expected linear scaling for the radiative limit without instabilities.

This factor 1/50 thus seems to represent a kind of  fixed ``shear/mixing penalty'' for X-ray production in the limit of strongly radiative shocks.

\section{Discussion \& Future Outlook}
\label{sec:summary}

All the above 2D planar expansion simulations  were computed with a fixed spatial mesh of grid size $\Delta x = 0.0026 \,d$.
For the highest-density model, $\mdot = 100 = 2d/\ell_c$, this is roughly half the 1D cooling scale $\ell_c$, indicating that this densest model could be resolution limited. Nonetheless,  note that it does still give roughly the same factor 1/50 reduction in X-ray emission seen in the 2D laminar-collision case, for which the grid is set to provide a quite high resolution of this cooling length, $\Delta x = 
 0.005\, \ell_c$.  
Additional planar expansion simulations done at a factor 2 lower resolution still give good general agreement to the ${\bar L}_X$ plotted in figure \ref{fig:masterPlot}, except that for the two highest-density models, with $\mdot =$ 56 and 100, there is a precipitous drop in X-ray emission. Since now the grid zones with  $\Delta x > \ell_c$ are no longer adequate to resolve the cooling length, even in the finger tips and troughs with compressive shocks the numerical hydrodynamics using the \citet{Tow2009} exact integration scheme for radiative cooling now simply jumps to the fully cooled post-shock equilibrium, without showing any of the intermediate shock heating.

Such resolution issues are an inherent challenge for any numerical effort to examine the effects of thin-shell instability on X-ray emission. As emphasized by \citet{ParPit2010}, there are many associated numerical effects, for example that of ``numerical conduction'', that can reduce and soften the associated X-ray emission. Our study here has not included any specialized attempts to mitigate such effects, but in the 2D planar advection model here, the fixed spatial grid has a resolution of the cooling length that is comparable to the adaptive mesh refinement models by \citet{ParPit2010}. Moreover, the vertical advection makes the vertical structure a convenient proxy for the time-dependent evolution. This illustrates how the thin-shell instability transforms  a laminar flow compression into an extensive shear structure, with associated reduction in X-ray emission that seems {\em not}  just a numerical effect, but grounded in the fundamentally different shear vs.\ compressive flow structure.  By using a systematic parameter study in the cooling efficiency associated with the mass source rate $\mdot$, the planar divergence models allow us to examine how the reduction of X-rays depends on the relative strength of adiabatic vs.\ radiative cooling.

A key result is the indication here that this thin-shell instability transition to complex structure might simply lead to a fixed reduction in X-ray emission. But further work will be needed to determine the origin of this reduction, and how it may be affected by either numerical or physical parameters.  For example, the factor 1/50 found here is roughly comparable to the pre-shock vs.\ post-shock sound speed, or the inverse Mach number. Future work should thus explore how this  factor depends on the assumed input/floor temperature $T_o$, or the shock inflow speed $ v_o$. In principle, this can be done within the simple laminar flow model, using perhaps adaptive mesh refinement to further increase the effective resolution. Such adaptive mesh methods would also be particularly useful in testing the inferred linear trend of X-rays at high $\mdot$.

Apart from this study of how colliding wind X-rays scale with cooling efficiency, it will be of interest to apply the insights here towards understanding the scaling of X-rays in other contexts, such as from the embedded wind shocks arising from the intrinsic instability of radiative driving of hot-star winds. In particular, \citet{OwoSun2013} have argued that the observed linear scaling between X-ray and stellar bolometric luminosity ($L_X \sim L_{bol}$) in single O-stars might be explained if thin-shell mixing of such embedded shocks reduces their X-ray emission by some power -- the ``mixing exponent'' $m \approx 0.4$ -- of the mass-loss rate.
This is distinct from the nearly constant X-ray reduction factor found here in the context of direct flow collision, and so future work should explore how the scaling of any X-ray reduction might depend on the specific geometrical and physical context of the shock production{, e.g. effect of a large-scale or turbulent magnetic field \citep{HeiSly2007}}.

Indeed, even within the context of CWB's, there are additional effects not considered here that could significantly alter how thin-shell instability affects X-ray production. For example, how will the factor 1/50 reduction found in these 2D models differ in a more realistic 3D wind collision? How might this be affected by the shear in an interaction between winds with different speed, on in the bow-shaped interaction of winds with differing momenta? Even in the planar interaction of winds with equal momenta, but unequal speed and density, the cooling parameter will be different on each side of the interaction front. How will this affect thin-shell structure and the reduction of X-ray emission? Clearly, there are many remaining issues for understanding the X-ray properties of such CWB's. But hopefully the current 2D study of thin-shell effects can form a good basis for addressing such more complex cases.

\section*{Acknowledgements}
This work was carried out with partial support by NASA ATP Grant  NNX11AC40G  
 to the University of Delaware.
 A.uD. acknowledges support from  NASA {\it Chandra} theory grant to Pennsylvania State University -- Worthington Scranton.
We thank D.~Cohen, J.~Sundqvist, V.~Petit and R.~Townsend for many helpful discussions.
We especially  thank C.M.P.~Russell for providing us tabulations of the radiative emission function $\Lambda (E,T)$ through his application of the APEC model in XSPEC. We also acknowledge use of the {VH-1} hydrodynamics code, developed by J.~Blondin and collaborators.

\bibliographystyle{mn2e}
\bibliography{thinshell}


\end{document}